\documentclass[aps,prd,preprint]{revtex4-2}
\makeatletter
\def\frontmatter@abstract@produce{%
  \par
  \addvspace{\frontmatter@preabstractspace}%
   \begingroup
    \dimen@\baselineskip
    \setbox\z@\vtop{\unvcopy\absbox}%
    \advance\dimen@-\ht\z@\advance\dimen@-\prevdepth
    \@ifdim{\dimen@>\z@}{\vskip\dimen@}{}%
   \endgroup
   \begingroup
    \prep@absbox
    \unvbox\absbox
    \post@absbox
   \endgroup
  \@ifx{\@empty\mini@notes}{}{\mini@notes\par}%
  \addvspace\frontmatter@postabstractspace
}%
\makeatother  

\pdfoutput=1
\usepackage{graphicx}
\usepackage{dcolumn}
\usepackage{bm}
\usepackage[hidelinks,colorlinks=true,allcolors=blue]{hyperref}
\usepackage{orcidlink}     

\usepackage{color}
\usepackage{slashed}
\usepackage[utf8]{inputenc}
\usepackage{amsmath,amssymb,multirow,graphicx, subfigure}
\usepackage{amsthm}
\usepackage{braket}
\usepackage[T1]{fontenc}
\usepackage{enumitem}
\usepackage[autostyle=true]{csquotes} 
\newcommand{\nc}{N_c}
\newcommand{\nf}{N_f}
\newcommand{\td}{T_{de}}
\newcommand{\tc}{T_\chi}
\newcommand{\tch}{T_\chi}
\newcommand{\tchi}{T_\chi}
\newcommand{\sul}{SU(N_f)_L}
\newcommand{\sur}{SU(N_f)_R}
\newcommand{\suv}{SU(N_f)_V}
\newcommand{\topq}{{\cal Q}}
\newcommand{\trace}{{\rm tr}\, }

\begin{document}

\title{Fractional anomalous determinants and the chiral phase transition}

\author{Robert D. Pisarski \orcidlink{0000-0002-7862-4759}}
\email{rob.pisarski@gmail.com}
\affiliation{Department of Physics, Brookhaven National Laboratory, Upton, NY 11973-5000}

\date{\today}

\begin{abstract}
At high temperature instantons form a dilute gas, so
in QCD-like theories the breaking of the anomalous $U(1)_A$ symmetry is given by integral powers of
the anomalous determinant, $\sim (\det \Phi)^{\topq}$, where
$\Phi\sim \overline{q}_L q_R$ is bilinear in the quark fields,
and with untwisted boundary conditions, the topological charge,
$\topq$, is an integer. 
A syncretic model is constructed,
which is manifestly "beyond Landau".  In the chiral limit,
at temperatures above the chiral phase transition, $T > \tchi$, only
integral powers of the anomalous determinant appear.
Below $\tchi$, following 't Hooft {\it et al.} I assume that the topological
charge $\topq$ is fractional, as an integer times $1/\nc$,
where $\nc$ is the number of colors.  I suggest that
consequently, fractional powers of the anomalous determinant appear in the chiral effective Lagrangian.  For $\nf$ degenerate flavors,
this generalizes the Witten-Veneziano term,
valid for small $\nf/\nc$, to arbitrary $\nf/\nc$.
In this model the chiral phase transition is
generically of second order.  The two exceptions are for one flavor, where it is probably crossover,
and three flavors, where it could well be weakly first order.  
This can be tested in lattice QCD with $2+1$ flavors by comparing the
(known) temperature dependence of the 
difference of the $\pi^a$ and $a_0^a$ propagators,
to the chiral condensate of the strange quark, between $\tchi$ and $\sim 2 \, \tchi$. Analogous measurements are possible for one to four degenerate flavors about $\tchi$.  Lastly, I propose an operator
for baryon number in the symmetric phase.
\end{abstract}. 

\maketitle

\section{Introduction}
\label{sec:intro}

The axial anomaly of Adler, Bell, and Jackiw is one of the
great surprises in quantum field theory \cite{Adler:1969gk,Bell:1969ts}.
It profoundly affects both condensed matter 
\cite{Armitage:2018rmp,Pettes:2025anomalies}
and high energy
\cite{Schafer-Nameki:2023jdn,Shao:2023gho}
physics, and    is deserving of the highest recognition.

In QCD, because of the anomaly, the divergence of the 
isosinglet current is proportional to the topological charge density
for the gluons.  On a closed manifold,
the integral of the topological charge density, $\topq$, is
an integer \cite{Belavin:1975fg}, so it is not surprising that 
the coefficient of the anomaly is exact at one loop order.

The anomaly is why the isosinglet meson, the
$\eta'$, is not a pseudo-Goldstone boson, like  
the octet of the $\pi$'s, $K$'s, and the $\eta$.  
Similarly, as for the $\eta'$ with spin zero, for mesons
with higher spin
the anomaly also splits the mass of the isosinglet "heterochiral" mesons from that of their octet partners \cite{Giacosa:2017pos,Giacosa:2025dvu}.

The anomaly is also central to the order of the chiral
phase transition, at a temperature $\tchi$.  Following Kobayashi, Maskawa, and
't Hooft \cite{Kobayashi:1970ji,Kobayashi:1971qz,Kobayashi:1973fr,Maskawa:1974,tHooft:1976rip,tHooft:1976snw}, anomalous terms in the effective Lagrangian
involve the determinant of the chiral field, $\sim \det \Phi$,
where $\Phi \sim \overline{q}_L q_R$. 
At high temperature (or density) 
\cite{Gross:1980br,Pisarski:1983ms,tHooft:1986ooh,Fejos:2014qga,Pisarski:2019upw,Fejos:2022mso,Bernhardt:2023hpr,Fejos:2024bgl,Pisarski:2024esv,Giacosa:2024orp,Fejos:2026tyd}, semi-classical calculations are reliable, and 
anomalous terms in the effective Lagrangian are integral
powers of the anomalous determinant.  Since for three flavors $\det \Phi$
is a cubic operator, if it dominates about $\tchi$, in the chiral
limit the transition is ---
{\it inescapably} --- of first order \cite{Pisarski:1983ms}.

Nevertheless, recently numerical simulations on the lattice
find {\it no} evidence for a first order transition.  By analytically
continuing in the number of flavors, 
Cuteri, Philipsen, and Sciarra \cite{Cuteri:2021ikv} assert that the chiral transition is of second order
for three massless flavors.  
Similarly, direct computation for three light flavors 
find no sign of a first order transition 
\cite{Cuteri:2021ikv,Dini:2021hug,DAmbrosio:2025imc,Zhang:2026zpa,Ding:2026blh}.
While a (very) weakly first order transition could still emerge, 
this region is much smaller than expected from effective models \cite{Pisarski:2016ixt,Resch:2017vjs,Giacosa:2024orp,Giacosa:2025dvu}.
This is mystifying, because in vacuum the $\eta'$ is much heavier than
the pseudo-Goldstone bosons in the octet.

In this paper I propose a syncretic model of the anomalous terms in the effective
Lagrangian for the chiral phase transition. The model is syncretic because the
anomalous terms differ below and above $\tchi$: below $\tchi$,
fractional powers of the anomalous determinant enter, $\sim (\det \Phi)^{1/\nc}$, where $\nc$ is the number of colors;
above $\tchi$, only integral powers appear.  That the anomalous terms in an effective Lagrangian 
differ below and above $\tchi$ is only possible if the transition is "beyond Landau", contrary to the usual paradigm of Landau, Wilson, Fisher, and Kadanoff
\cite{Senthil:2003eed,Senthil:2004fuw,Senthil:2023vqd,Chen:2025generalized}.
Previous analysis by Wilczek and I suggested that the chiral transition is of first
order for three or more flavors \cite{Pisarski:1983ms}.  In the
present model, 
the chiral phase transition is generically of 
second order, except for one and three flavors.
For one flavor, it is probably crossover, while for three flavors,
it is weakly first order, possibly very weakly.  

Previously, Chen, Cherman and I (CCP) considered
the phase transitions of QCD-like theories in the plane of temperature
and an imaginary baryon chemical potential, 
$\theta_B$ \cite{Chen:2026jla}.  We showed that in the $T - \theta_B$
plane, the chiral
phase transition probably has a critical line whose critical
exponents change with $\theta_B$.
This is then the first example in which the chiral phase transition of
QCD exhibits critical behavior which is beyond Landau.

The fractional powers of the anomalous determinant are 
due to fractional instantons, with
topological charge $\topq = 1/\nc$ times an integer.  
While on
a closed manifold such as the four sphere 
the topological charge
is an integer, 't Hooft first showed that the topological
charge is fractional on manifolds with 
$Z(\nc)$ twisted boundary conditions \cite{tHooft:1981nnx}.  
Notably, Lee, Lu, Kraan, and van Baal
showed that for $SU(\nc)$ instantons at nonzero temperature, they can be characterized as having $\nc$ constitutents, each    
with topological charge $1/\nc$ \cite{Lee:1998bb,Kraan:1998pm,Kraan:1998sn}.
See, also, the analysis of Nair and I \cite{Nair:2022yqi}.

As discussed by Gonzalez-Arroyo \cite{Gonzalez-Arroyo:2023kqv},
numerical simulations of pure $SU(\nc)$ gauge theories
on the lattice find evidence
for fractional topological charge. 
This especially clear from
the $\theta$-dependence for $\nc =3$, $4$, and $6$,
Fig. (8) of Bonanno, Bonati, and D'Elia \cite{Bonanno:2020hht}.

A pure gauge theory has a deconfining phase
transition at a temperature $\td \approx 270$~MeV \cite{Boyd:1996bx,Lucini:2005vg}. At large $\nc$, lattice QCD finds that
the topological susceptibility is large and approximately independent
of $T$ when $T \leq \td$, and very small, falling sharply with $T$, above $\td$
\cite{Vicari:2008jw,Ce:2016awn,Borsanyi:2022fub,Bonanno:2023hhp,Bonanno:2025abc}.
This is natural if fractional instantons dominate below $\td$, and
integrally charged instantons dominate above.

In contrast, for QCD with $2+1$ light flavors, there is a crossover for
the chiral transition at a much lower temperature, at $\tch \approx 155$~MeV \cite{HotQCD:2018pds,Borsanyi:2020fev}.  I suggest that with massless quarks, fluctuations with topological charge are
dominated by fractional instantons below $\tch$, and integrally
charged instantons above $\tchi$.
Since $\tch$ is about half of $\td$, at first sight this seems
odd, since at least semi-classically all topologically nontrivial configurations
are purely gluonic.  In the strict chiral limit, 
however, I suggest that above $\tch$, the zero modes of massless quarks bind
fractional instantons into instantons with integral charge.   
This doesn't happen in the low temperature phase, because there are no
zero modes if the quarks have a dynamically generated mass.  
Consequently, in the chiral limit $\tchi$, and the temperature at
which the $U(1)_A$ is approximately restored, should coincide.  
For two flavors in the continuum limit, there is numerical evidence for
this by Chiu and Hsieh \cite{Chiu:2026sxy,Chiu:2026upk}.

A useful limit is that of
"femto"-slabs, where one (or more) spatial dimensions are so small that by asymptotic freedom the gauge
coupling is also small, and a semi-classical 
analysis is valid.
On a femto-slab, fractional instantons appear because in three 
dimensions, topological charge is carried by monopole-instantons;
see the review by Poppitz \cite{Poppitz:2021cxe}.

My suggestion is inspired by two examples.  First, 
the 't Hooft limit of large $\nc$, where 
$\nc \rightarrow \infty$ and $\nf/\nc \rightarrow 0$ \cite{tHooft:1973alw,Witten:1979kh,Lucini:2012gg,Bali:2013kia}.
Instead of the linear field $\Phi$, in vacuum it is customary
to use the non-linear field $U$, which, unlike $\Phi$, is a unitary matrix.
In the 't Hooft limit, the only anomalous term in the chiral
effective Lagrangian is a mass term for the $\eta'$, which is
light, $m_{\eta'}^2 \sim \nf/\nc$.  Witten
and Veneziano showed that in terms of the non-linear field $U$,
the chirally invariant form for the 
$\eta'$ mass is non-polynomial and multi-valued, $\sim (\nf/\nc)|\log \det U|^2$
\cite{Witten:1979vv,Witten:1980sp,Veneziano:1979ec,DiVecchia:1980yfw}.

An obvious question is then how the Witten-Veneziano term
generalizes from small $\nf/\nc$, where the $\eta'$ is light, to large $\nf/\nc$,
where the $\eta'$ is heavy, as for QCD.
I find it essential to consider the Veneziano limit
\cite{Veneziano:1974fa,Veneziano:1976wm}, where both $\nf$ and $\nc$ are large,
and the $\eta'$ is heavy.

The second limit is that of femto-slabs, where the zero modes "fractionalize"
\cite{Cherman:2016hcd,Tanizaki:2022ngt}.  On a femto-torus,
fractional instantons are $Z(\nc)$ vortices, and
fractional powers of anomalous determinants, 
$ \sim (\det U)^{1/\nc}$, arise \cite{Hayashi:2024qkm}.  

The matrix $U$ is a product of a unitary matrix, involving the pions
and a phase for the $\eta'$ meson.  Any unitary matrix has determinant
one, so terms such as $ \sim (\det U)^{1/\nc}$ are 
$\sim {\rm e}^{i \eta'/(f_\pi \nc)}$, and so are innocuous.
This is not true for fractional powers of  
a linear field, $\Phi$, as $\sim (\det \Phi)^{1/\nc}$, as
such fractional powers 
are not standard in the usual Wilsonian approach to the renormalization group.
At least for the chiral phase transition,
and perhaps other examples --- such as in Weyl semi-metals
\cite{Zyuzin:2012tv,Goswami:2013PRB,Wang:2013PRB,Gooth:2019lmj,Mottola:2023emy} --- the axial anomaly forces their appearance.  Indeed,
the reason that that I suggest that fractional powers 
of the anomalous determinant only 
appear in the broken phase, for $T < \tchi$, 
is simply because they seem nonsensical in the symmetric phase.

I propose several measurements which can be done on the lattice
to test my proposal.  
I first discuss how twisted boundary conditions
with overlap fermions can provide direct evidence.
Since $Z(\nc)$ twisted boundary conditions with overlap fermions are necessary, though, this is not easy.

The second method is indirect, but is a minor extension of known results.
Many groups have computed the 
difference of the two point functions at zero momentum
for the $\pi^a$ and $a_0^a$, as test of when the axial $U(1)_A$ symmetry
is approximately restored
\cite{Bazavov:2012qja,Buchoff:2013nra,Bhattacharya:2014ara,Aoki:2015qwa,Bonati:2015vqz,Dick:2015twa,Borsanyi:2016ksw,Tomiya:2016jwr,Bonati:2018blm,Kotov:2019dby,Ding:2020xlj,Aoki:2020noz,Borsanyi:2020fev,Kaczmarek:2020sif,Aoki:2021qws,Kaczmarek:2021ser,Lahiri:2021lrk,Borsanyi:2022fub,Kaczmarek:2023bxb,Alexandru:2024tel,Borsanyi:2025lim,Kotov:2025ilm,Aarts:2026kpq,Chiu:2026sxy,Chiu:2026upk}.
The only other quantity needed is the value 
of the (renormalized) chiral condensate for the strange
quark.  

In this paper I give the form of chiral effective Lagrangians
for one to four degnerate flavors.  
The case of non-degenerate flavors is non-trivial
and treated separately 
\cite{Giacosa:2026,Kovacs:2026,Debnath:2026}.
Results from the lattice will directly
allow the parameters
of these effective Lagrangians to be fit.  
For QCD with $2+1$ flavors, 
if $\tchi$ is the temperature for the chiral crossover, it is best to measure
between $\tchi$ and $2 \, \tchi$, where the chiral condensate for the 
strange quark evaporates. This will be most useful for phenomenology,
and in particular, for computing the phase diagram of QCD in the
plane of temperature and chemical potential  \cite{Debnath:2026,Nogradi:2026}.
For $\nf$ symmetric, light flavors,
$\nf = 1 \rightarrow 4$, it is best to measure about $\tchi$.

Lastly, I suggest an operator for baryon number in the symmetric
phase of a linear sigma mode, which allows one to connect the present
analysis to that of CCP \cite{Chen:2026jla}.

\section{The large $\nc$ limit of 't Hooft}
\label{sec:tHooft}

I begin by reviewing the 't Hooft limit of large $\nc$,
where the number of massless flavor in the fundamental
representation, $\nf$, is held
fixed as the number of colors, $\nc \rightarrow \infty$
\cite{tHooft:1973alw,Witten:1979kh,Lucini:2012gg,Bali:2013kia}.
In this limit, precise statements can be made about
the anomalous terms in an effective Lagrangian
\cite{Witten:1979vv,Witten:1980sp,Veneziano:1979ec,DiVecchia:1980yfw}.
In the next section I consider the Veneziano
limit, when both $\nf$ and $\nc$ are large, but
held in fixed ratio.   

As a local, non-Abelian gauge theory, only color singlet states
are physical ("confinement"), comprising glueballs, mesons, and baryons.
The masses of glueballs and mesons are $\sim \nc^0 \sim 1$, with
baryon masses $\sim \nc$.

Then at infinite $\nc$ the theory is free.  Two point functions of gluonic fields
fields are $\sim \nc^2$, while those of quarks are
$\sim \nc$.  Assuming the masses of
order one, the form factor to create a glueball is $\sim \nc$, so
the three and four glueball couplings are $\sim 1/\nc$ and
$\sim 1/\nc^2$;
the three and four meson couplings are $\sim 1/\sqrt{\nc}$
$\sim 1/\nc$, {\it etc.}  Baryon masses
are $\sim \nc$, so their couplings, and those to mesons, are large \cite{Witten:1979kh}.   

In the vacuum, the chiral symmetry is spontaneously broken as \cite{Chen:2026jla}
\begin{equation}
\frac{U(1)_B \times \sul \times \sur }{Z(\nf)}
\; \Rightarrow \; \frac{U(1)_B \times \suv}{Z(\nf)} \; .
\label{breaking_symmetry}
\end{equation}
This pattern can be proven at infinite $\nc$ \cite{Coleman:1980mx}, and agrees
with hadronic phenomenology.  The classical symmetry of $U(1)_A$
is not included,
as that is broken quantum mechanically
through the axial anomaly \cite{Adler:1969gk,Bell:1969ts}. Even so,
the central theme of this paper is the way in which the breaking of axial $U(1)_A$ symmetry manifests itself
in effective Lagrangians for the chiral phase transition.

Since the theory is free at infinite $\nc$, the thermodynamics at $\mu =0$
follows directly.  The confined phase is free until a temperature $\td$,
when it undergoes a first order transition to a deconfined phase.  The
pressure is $\sim \nc^0$ for $T \leq \td$, and $\sim \nc^2$ for $T > \td$.
Since the gluons dominate the free energy, the temperature for
the restoration of chiral symmetry, $\tchi$, surely happens at the same
time, with $\tchi = \td$.  This is valid up to small corrections
$\sim \nf/\nc$, so the pressure is $\sim \nf \nc^0$ at $T < \td$,
{\it etc.}.

To understand the quantum mechanical breaking of the $U(1)_A$ symmetry,  start with the vacuum energy at
nonzero $\theta$, which is the CP-odd angle in a gauge
theory \cite{Bonanno:2026zzf}.  Then the vacuum energy at nonzero $\theta$
and $T \neq 0$ can be expanded about $\theta = 0$ as
\begin{equation}
E(\theta,T) - E(0,0) 
= \, 
\chi^{{\rm pure} \, {\rm glue}}_2 
\;\; \frac{ \theta^2}{2}\left( 1 + b_2 \, \theta^2 + \ldots \right) \; .
\label{top_suscep_conf}
\end{equation}
The topological susceptibility $\chi^{{\rm pure} \, {\rm glue}}_2 \sim \nc^0$ is of order one and independent of $T$ in the confined
phase, $T < \td$, as $\nc = \infty$ is a free
field theory of mesons and glueballs.  This is confirmed by lattice simulations at large $\nc$ \cite{DelDebbio:2004vxo,Bonati:2013tt,Bonati:2016tvi,Bonati:2018rfg,Bonanno:2019nze,Bonanno:2020hht,Bonanno:2023hhp,Bonanno:2025abc}.

In contrast, at asymptotically high temperature this can be computed
semi-classically, where it is dominated by instantons
\cite{Belavin:1975fg,tHooft:1976snw,tHooft:1976rip,Gross:1980br}.
Instantons carry a topological charge, 
$\topq$, with an action $= (8 \pi^2/g^2) |\topq|$,
where $g$ is the gauge coupling constant.  Since $g^2 \nc = \widetilde{g}^{\, 2}$
is held fixed as $\nc \rightarrow \infty$, the instanton action is
$= (8 \pi^2/\widetilde{g}^{\, 2}) \nc$, and so in the path integral 
their contribution is exponentially suppressed, $\sim \exp( -(8 \pi^2/\widetilde{g}^{\, 2}) \nc)$.

The most plausible explanation is the existence of fractional instantons,
with topological charge $\topq = \pm 1/\nc$ times an integer.  
Then their action is proportional to powers of $\exp(- 8 \pi^2/\widetilde{g}^{\, 2})$, and
survives as $\nc \rightarrow \infty$ \cite{tHooft:1981nnx,Gonzalez-Arroyo:2023kqv,Nair:2022yqi,Poppitz:2021cxe}.

Witten and Veneziano 
\cite{Witten:1979vv,Witten:1980sp,Veneziano:1979ec,DiVecchia:1980yfw}
showed how topological fluctuations show up
in the mesonic effective Lagrangian in the 't Hooft
limit of large $\nc$.  If the quarks are massless,
the $\theta$-angle can be eliminated by an axial rotation.  
Because the contribution of quarks is suppressed at large $\nc$,
the anomaly only shows up as a small mass for the isosinglet $\eta'$ meson, with
\begin{equation}
m^2_{\eta '} = \frac{4 \nf}{f_\pi^2} \;
\chi^{{\rm pure} \,{\rm glue}}_2
 \sim \frac{\nf}{\nc} \;  ;
\label{etaprime_mass}
\end{equation}
$f_\pi \sim \sqrt{\nc}$ is the pion decay constant,
and the topological susceptibility 
$\chi^{{\rm pure} \,{\rm glue}}_2 \sim \nc^0$ is of
order one.  Like $\chi^{{\rm pure} \,{\rm glue}}_2$,
the mass of the $\eta'$ is independent
of $T$ in the confined phase, $T < \tchi = \td$.

An effective theory of the Goldstone bosons,
and one almost Goldstone boson, the $\eta'$, can be constructed  
by using a nonlinear representation, with 
$U(x)$ a $U(\nf)$
matrix that transforms under $\sul \times \sur \times U(1)_A$ as
\begin{equation}
U(x) \rightarrow {\rm e}^{i \alpha_A} \, \Omega_L^\dagger \, U(x) \, \Omega_R \; .
\label{U_transform}
\end{equation}
Here $\Omega_L$ and $\Omega_R$ are $\sul$ and $\sur$ rotations, while $\exp( i\alpha_A)$ is a rotation in axial $U(1)_A$.

The Lagrangian starts with a kinetic term,
\begin{equation}
{\cal L}_{NL} = f_\pi^2 \, {\rm tr} \, |\partial_\mu U(x)|^2 + \ldots
\label{nonlinear_L}
\end{equation}
plus terms with higher derivatives.  $U(x)$ 
is a matrix in $U(\nf)$, so we
can write it as a product of an
element of $SU(\nf)_V$, $U_V(x)$, times
an element of $U(1)_A$.  The latter is a phase
proportional to the $\eta'(x)$ field,
\begin{equation}
U(x) = {\rm e}^{i , \eta'(x)\, t^0/f_\pi}
\; U_V(x) \; ; \;
U_V(x) = {\rm e}^{ i \, \pi^a(x) \, t^a/f_\pi}
\; .
\label{def_U}
\end{equation}
where the $t^a$, $a = 1\ldots(\nf^2 -1)$, are the generators of $SU(\nf)$,
$\trace t^a t^b = \delta^{a b}/2$,
and $t^0 = {\bf 1}_{\nf}/\sqrt{2 \nf}$ that for $U(1)$.  
$U_V$ is an element of $SU(\nf)_V$, 
so $U_V^\dagger U_V = {\bf 1}_{\nf}$, and
$\det U_V = 1$.

In the 't Hooft large $\nc$ limit, Witten and Veneziano showed
that the only anomalous term is just
a mass for the $\eta'$.  The chirally invariant form
is non-polynomial and multi-valued in $\det U$
\cite{Witten:1979vv,Witten:1980sp,Veneziano:1979ec,DiVecchia:1980yfw},
\begin{equation}
{\cal L}_{WV} = f_\pi^2 \, m_{\eta'}^2 \, | \log \det U(x)|^2  =
 m_{\eta'}^2 \; \eta'(x)^2 \; .
\label{witten-venez}
\end{equation}
This is valid in vacuum, and since the theory is free in the confined phase,
is unchanged for all $T < \td$.

Witten emphasized that in the 't Hooft limit, there is a 
subtle property of the
energy of the vacuum which follows from Eq. \eqref{witten-venez}.  
As a function of $\nc$, 
\begin{equation}
E(\theta,\nc) - E(0,\nc) = \nc^2 \; \widetilde{E}_{tH} \left( \frac{\theta}{\nc} \right)
\left( 1 + O\left(\frac{1}{\nc}\right) \right) \; .
\label{theta_nc_tHooft}
\end{equation}
This arises because the logarithm of $\det U$ in the Witten-Veneziano term
is multi-valued, so for large $\nc$, there are $\sim \nc$ metastable
vacua.  Since the vacuum energy is periodic in 
$\theta \rightarrow \theta + 2 \pi$, a function of $\theta/\nc$ 
must be multi-valued, or the period would instead be $2 \pi \nc$.  This property 
is used in deriving the generalization of the Witten-Veneziano term
to the Veneziano limit of large $\nc$.

In the deconfined phase, $T > \td$,
the density of instantons is exponentially small, $\sim \exp(- (8 \pi^2/\widetilde{g}^{\, 2}) \nc)$,
and thus so are all anomalous terms in the chiral effective
Lagrangian.  Even so, a semi-classical computation applies,
and it is illuminating to outline what these exponentially small
terms are.  

The deconfined phase is chirally symmetric, so I introduce a linear
field $\Phi(x)$, as a $\nf \times \nf$ complex valued matrix.
This has $ 2 \nf^2$ degrees of freedom, which includes all the
lowest mesons with $J^P = 0^\pm$ in the symmetric phase.
This is in contrast to the nonlinear model, which only includes the Goldstone
bosons.  $\Phi(x)$ transforms under global flavor
rotations as $U(x)$ does in Eq. \eqref{U_transform}.  The effective Lagrangian
is composed of a piece which is symmetric under axial $U(1)_A$ rotations,
\begin{equation}
  \mathcal{L}_{U(1)_A \; {\rm sym}} =
  \trace\left|\partial_\mu \Phi\right|^2 
  + m^2 \, \trace \Phi^\dagger \Phi 
  + \lambda_1\, (\trace\, \Phi^\dagger\Phi )^2\,
  + \lambda_2 \, \trace (\Phi^\dagger\Phi )^2  + \ldots
  \label{eq:eff_lag_ua1_sym}
\end{equation}
neglecting terms $\sim \Phi^6$, {\it etc.}.   Since these
terms involve traces of powers of 
$\Phi^\dagger \Phi$, they are invariant under $U(1)_A$.
Terms which are not invariant under $U(1)_A$ must involve
powers of the anomalous determinant.  In the deconfined phase,
these are all integral powers of $\det \Phi$,
\begin{equation}
 {\cal L}_{{\rm int.}\, \topq} = \xi_1 \, (\det\Phi + \det \Phi^\dagger)  
  + \xi_1^{1} \, ( \trace \Phi^\dagger \Phi ) \, 
  (\det\Phi + \det \Phi^\dagger)
  + \xi_2 ( (\det\Phi)^2 + (\det \Phi^\dagger)^2 ) + \ldots
\label{eq:eff_lag_ua1_vio}
\end{equation}
The subscript on the Lagrangian
denotes that these are due to terms with integral topological charge, $\topq$.
As $\det \Phi \sim \Phi^{\nf}$, 
for three flavors these are of third, fifth, and sixth order, respectively \cite{Pisarski:1983ms,Pisarski:2019upw,Pisarski:2024esv,Giacosa:2024orp,
Giacosa:2025dvu}.  There are similar anomalous
couplings for
heterochiral mesons with higher spin
\cite{Giacosa:2017pos}
involving polydeterminants \cite{Giacosa:2025dvu}.

These terms are generated by zero modes for the quarks in topologically
non-trivial fields.  An instanton with topological charge $|\topq|$
has $\nf |\topq|$ zero modes for the left and right-handed quarks
or anti-quarks, where the sign of $\topq$ determines which are quarks,
and which are anti-quarks.  While instantons do not contribute to the
partition function, they do once sources with non-zero chirality are
introduced.  In this way, the first two terms, $\sim \det\Phi$, are generated by topologically
nontrivial configurations with $\topq = \pm 1$
\cite{Belavin:1975fg,tHooft:1976rip,tHooft:1976snw,tHooft:1986ooh}, and the last, $(\det \Phi)^2$, is
produced by configurations with $\topq = \pm 2$
\cite{Pisarski:2019upw}, {\it etc.}  Consequently, in the chirally
symmetric phase the vacuum energy is single valued, and the periodicity
in $\theta \rightarrow \theta + 2 \pi$ is direct.

For both Eqs. \eqref{nonlinear_L} and \eqref{eq:eff_lag_ua1_vio}, 
the $\theta$-dependence is included merely by replacing $U \rightarrow
{\rm e}^{i \theta} U$, and $\Phi \rightarrow {\rm e}^{i \theta} \Phi$.
Doing so, it is immediate that the $\theta$-dependence can
be eliminated by a rotation in $\eta'$.  Thus the $\theta$-dependence
arises uniquely from mass terms.  In the phase with chiral
symmetry breaking, to leading
order in the mass matrix $M$ this is
\begin{equation}
{\cal L}_M =  \trace \left(  {\rm e}^{i \theta}\,  U(x)  \, M + {\rm e}^{- i \theta}\,  U^\dagger(x)\,  M^\dagger \right) \; ,
\label{mass_eq}
\end{equation}
In the chirally symmetric phase one just replaces $U(x)$ by $\Phi(x)$;
I assume that $M$ is a diagonal matrix.

\section{The large $\nc$ limit of Veneziano}
\label{sec:veneziano}

The Veneziano limit is where both $\nf$ and $\nc$ are large, 
with $\nf/\nc$ held fixed.  Since confinement holds, all states
are color singlets, with interactions suppressed by powers of 
$1/\sqrt{\nc}$, {\it etc.}  However, this can be compensated by
factors of $\nf$, so generally the theory is
strongly interesting until asymptotically high temperature.

Numerical simulations with large $\nf$ and $\nc$ are sparse \cite{DeGrand:2023hzz,DeGrand:2021zjw}, while there are many
for QCD, with $\nf = 2+1$ and $\nc = 3$.  
Thus I blithely assume that QCD is close to the Veneziano limit.
Obviously $\nf/\nc$
cannot be too large: for $\nc = 3$, in vacuum chiral symmetry is not
broken if $\nf > \nf^*$, where  perhaps $7 \leq \nf^* \leq 8$
\cite{Pastor-Gutierrez:2026ujh}.
Presumably $\nf^*/\nc \approx 2 - 3$ at large $\nc$.
In any case, $\nf/\nc < \frac{11}{2}$ so that asymptotic freedom is not lost.

Since the theory is strongly interacting,
the pressure is no longer an order parameter, since
there are $\sim \nf^2$ and $\sim \nf \nc$ degrees of freedom for the quarks, and $\sim \nc^2$ for the gluons.
In QCD, with $\nf = \nc = 3$, the pressure is still small at low
temperature \cite{HotQCD:2018pds,Borsanyi:2020fev}, because numerically the number of degrees of freedom for pions and kaons, with spin zero,
is much less than those for quarks, with spin $\frac{1}{2}$,
and gluons, with spin $1$.

As mentioned previously, for three colors the pure glue theory has 
a deconfining phase transition of first order at $\td \approx 270$~MeV
\cite{Boyd:1996bx,Lucini:2005vg}.
The overall mass scale is fixed by assuming the string tension
is the same as in QCD,which is a notable, but standard, assumption.

QCD has $2+1$ light flavors, and the first order deconfining transition is washed out by dynamical quarks.  Instead, there is a crossover, as observed from a peak in
the chiral susceptibility for light quarks
\cite{HotQCD:2018pds,Borsanyi:2020fev}, at 
$\tchi \sim 155 \pm 5$~MeV. 

Henceforth, by $\td$ I always mean the temperature deconfinement in the pure glue
theory.  Thus in QCD, there is about a factor of two difference between
$\tchi$ and $\td$.   As $\nf$ increases, $\tchi$ presumably decreases, until $\tchi =0$ at $\nf^*$.  

To simplify the discussion I henceforth assume that $\nf = \nc$, with
$\td \approx 2 \, \tchi$.
In the pure glue theory, the Polyakov loop jumps to about $0.5$ at $\td$,
more or less independent of $\nc$ \cite{Mykkanen:2012ri}; by
the Polyakov loop I always mean the renormalized loop in the
fundamental representation.  
With dynamical quarks the Polyakov loop is nonzero at any
$T \neq 0$.  Even so, in QCD the value of the Polyakov loop
is {\it tiny} at $\tchi$, only $\approx 0.1$, while its value
is near unity at $2 \, \tchi$ \cite{Petreczky:2015yta}.  Thus even with dynamical quarks,
near $\tchi$ is theory is close to a confined regime.  I
use the term "deconfinement" when the Polyakov loop is near one,
but stress that this term is only suggestive.
For this reason, the term is always written in quotes
\footnote{
Lattice simulations have measured the susceptibilities for both $\Phi$ and the Polyakov loop.  That for $\Phi$ exhibits true critical behavior.
That for non-critical fields, such as the Polyakov loop, do
exhibit peaks in their associated susceptibilities near $\tchi$,
but this is due entirely to their coupling to the critical field, $\Phi$.
This was first noted by Sasaki, Friman, and Redlich \cite{Sasaki:2006ww},
and is obvious in a chiral matrix model, see Eq. (111) and Fig. (9)
of Ref. \cite{Pisarski:2016ixt}.  Thus a peak in the loop susceptibility
near $\tchi$ does {\it not} indicate deconfinement \cite{Borsanyi:2025lim}.
}.

What happens to the behavior of topological structure in the vacuum? Whether or not there are many quarks about, as a configuration of pure
glue, the instanton action is still $\sim (8 \pi^2/\widetilde{g}^{\, 2}) \nc$, and so instantons
are exponentially suppressed.  Thus as in the 't Hooft large $\nc$
limit, at large temperature the topological susceptibility is
exponentially small, $\sim \exp(- (8 \pi^2/\widetilde{g}^{\, 2}) \nc)$
\footnote{
For massless quarks the topological susceptibility
vanishes, so I presume that the quarks masses are small,
but nonzero.  Alternately, one can compute the anomalous couplings of the chiral effective
Lagrangian, as these do not vanish in the chiral limit.}.

Indeed, numerical simulations of QCD support a large change
in the topological susceptibility above $\tch$
\cite{Luscher:2010ik,Schaefer:2010hu,Bonati:2013tt,ETM:2015bzg,Bonati:2015vqz,Petreczky:2016vrs,Frison:2016ivz,Borsanyi:2016ksw,Ce:2016awn,Taniguchi:2016tjc,Giusti:2018mdh,Bonati:2018blm,Bonanno:2019nze,Athenodorou:2022aay,Kotov:2025ilm}: 
from $\tc$ to $2 \tc$,
$\chi^{QCD}_2(T)^{1/4}$ falls by $\approx 2/7$ \cite{Petreczky:2016vrs}, so that by
$2 \, \tchi$, the topological susceptibility has fallen
by $\sim (2/7)^4$, or to about {\it one percent} of its value at $\tc$:
\begin{equation}
\chi^{QCD}_2(2 \, \tchi)\approx 0.01 \, \chi^{QCD}_2(\tc)
\label{ratio_top_suscep}
\end{equation}

In QCD, the lattice suggests that there is a dilute instanton
gas above $\approx 2 \, \tc$.  Since in QCD $g^2(T) \sim 1/\log(T)$, 
in the dilute gas regime
the topological susceptibility decreases as a power of $1/T$
\cite{Luscher:2010ik,Schaefer:2010hu,Bonati:2013tt,ETM:2015bzg,Bonati:2015vqz,Petreczky:2016vrs,Frison:2016ivz,Borsanyi:2016ksw,Ce:2016awn,Taniguchi:2016tjc,Giusti:2018mdh,Bonati:2018blm,Bonanno:2019nze,Athenodorou:2022aay,Kotov:2025ilm}.

Any effective Lagrangian must satisfy that it is independent of
$\theta$ for massless quarks.  Thus the $\theta$-dependence is controlled entirely by the mass term of Eq. \eqref{mass_eq}.  Hence 
\begin{equation}
\chi_2(T) \approx \trace \left( M \langle \Phi\rangle \right) \; ; \; T < 2 \tchi \; .
\label{semi_top_suscep}
\end{equation}
Once the renormalized chiral condensates are measured, matching this to the semi-classical
computation in a dilute instanton gas, valid
for $T > 2 \, \tchi$ \cite{tHooft:1976rip,tHooft:1986ooh,Pisarski:2019upw},
provides a real constraint on both. 
The instanton density has been computed only at
one loop order, so it would be worthwhile, albeit challenging,
to compute the instanton density to two loop order.

I take the model to be syncretic, so at low temperature fractional powers of the anomalous determinant contribute, while at high temperature, only integral powers do.  Note that since the strange quark
mass is not small, it is reasonable to assume that fractional instantons,
and thus fractional powers of the anomalous determinant,
persist as long as the chiral condensate for at least
one quark flavor --- in QCD that for the strange quark --- has a significant expectation value.

To describe both sides of the chiral phase
transition, instead of the non-linear field $U$, which only
includes the Goldstone bosons (and the $\eta'$), I use
a linear field, $\Phi$, as that includes
all of the degrees of freedom in the symmetric phase.

A first try for the anomalous terms
in the effective Lagrangian for $\Phi$ is just to replace $U$ by $\Phi$ in Eq. \eqref{witten-venez},
\begin{equation}
\frac{1}{\nc} \, | \log \det \Phi|^2 \; .
\label{wrong_gwv}
\end{equation}
This is unsatisfactory, though, since it diverges if $\det \Phi = 0$.
There are many matrices  where traces of powers of $\Phi^\dagger \Phi$ are 
nonzero, but $\det \Phi$ vanishes.  For example, take a matrix where the first diagonal element is nonzero, and all other elements vanish.

Useful insight can be gained from  
QCD-like theories in a femto-slab, where one or more more lengths is
femtoscale, of size $L$.  Then the QCD coupling constant,
$g^2(L) \ll 1$, and semi-classical methods can be used
\cite{Poppitz:2021cxe}.  Most illuminating are studies of
femto-slabs, by 
Cherman, Schaefer, and Unsal \cite{Cherman:2016hcd} and
by Tanizaki and Unsal
\cite{Tanizaki:2022ngt}, and on a
femto-torus, by Hayashi and Tanizaki
\cite{Hayashi:2024qkm}.  In these femto-slabs,
the dominant topologically nontrivial configurations are not instantons, with $\topq = \pm 1$,
but monopole-instantons, with $\topq = \pm 1/\nc$.  Further, they
exhibit a fractionalization of zero modes.

Hayashi and Tanizaki \cite{Hayashi:2024qkm}
showed that on a femto-torus, fractional instantons 
are $Z(\nc)$ vortices.  Because a quark in the fundamental representation
picks up a phase of ${\rm e}^{2 \pi i/\nc}$ as it encircles a
$Z(\nc)$ vortex, the anomalous terms in the chiral
effective Lagrangian are powers of
$\sim (\det U)^{1/\nc}$.  On a femto-torus the effective theory lies in
$1+1$ dimensions, so a $Z(\nc)$ vortex lives at a point, which
simplifies things.  In higher dimensions,
fractional anomalous determinants may still arise
from the interaction of quarks with $Z(\nc)$ vortices \cite{hidaka:2026},
although since $Z(\nc)$ vortices are then
extended objects, the derivation
is not trivial.

In the phase with broken chiral symmetry, 
a fractional power of $\sim (\det U)^{1/\nc}= 
\exp( i \eta'/(f_\pi \nc))$ is innocuous, but this is not true for a linear field, $\Phi$.  A first guess is just to
replace $U \rightarrow \Phi$,
\begin{equation}
(\det \Phi)^{1/\nc} + (\det \Phi^\dagger)^{1/\nc} \,
\label{first_guess}
\end{equation}
which at least is finite when $\det \Phi=0$.
Still, it cannot be quite right, as $(\det \Phi)^{1/\nc} \sim \Phi^{\nf/\nc}$
as $\Phi \rightarrow 0$.  Near a second order transition at $\tchi$, I
assume that only integral powers of $\Phi$ enter, as is standard
in the Wilsonian renormalization group.

Thus I impose two further conditions.  First,
that the $\eta'$ meson is massive at $\tchi$.  Second, that fractional powers of 
the determinant of
$\Phi^\dagger \Phi$ are introduced to make all terms well behaved as
$\Phi \rightarrow 0$.
For example, for the simplest term of Eq. \eqref{first_guess}, 
one might take
\begin{equation}
{\cal L}_{{\rm anom}}^1 = 
m_1^2 \; (\det \, \Phi^\dagger \Phi)^{1/\nf - 1/(2 \nc)}
\; \left( (\det \Phi)^{1/\nc} +  (\det \Phi^\dagger )^{1/\nc}\right) \; .
\label{gwv}
\end{equation}
The power of $1/\nf - 1/(2 \nc)$ for
$\det \, \Phi^\dagger \Phi$ is chosen so that
in all, ${\cal L}_{{\rm anom}}^1 \sim \Phi^2$ as $\Phi \rightarrow 0$.

In general, I propose that in the broken phase the complete anomalous Lagrangian is
\begin{equation}
{\cal L}_{{\rm anom} } = \sum^{\infty}_{j = 1}
{\cal F}_j(\Phi^\dagger \Phi) \; \left( (\det \Phi)^{j/\nc} +
(\det \Phi^\dagger)^{j/\nc} 
 \right) \; .
\label{general_fracQ}
\end{equation}
This is a sum over {\it all} instanton charges, fractional and integral.
It is not obvious what the functions ${\cal F}_j$ are.  
I assume that they are functions only of $\Phi^\dagger \Phi$, which is invariant
under $U(1)_A$, as then the terms $\sim (\det \Phi)^{j/\nc}$ have 
well defined charges under axial rotations.

Above $\tchi$ the anomalous terms must match onto those
in an instanton gas, so I require that in the strict chiral limit,
all terms with fractional powers of $\det \Phi$ vanish when $T > \tchi$,
\begin{equation}
{\cal F}_j = 0 \;\; ; \;\; j \neq n \; \nc \;\; ; \;\; T > \tchi \; ,
\label{Fs_above_tch}
\end{equation}
where $n$ is an integer.  Even so, I do {\it not} assume
that there is a dilute instanton gas above $\tchi$, only that $\topq$ is
an integer.

I also assume that for the terms with integral topological charge,
that the coefficients ${\cal F}_j$ are constants, with no peculiar functions of 
$\Phi^\dagger \Phi$.  This is the easiest way for
the terms with integral $\topq$ at $T \leq \tchi$ to match onto
those at $T > \tchi$.

I note that fractional powers of superfields
arise in supersymmetric models \cite{Veneziano:1982ah,Taylor:1982bp,Affleck:1983mk}.
There, however, holomorphy is useful, which is not possible here.

At large $\nc$, in vacuum I assume that $\langle \Phi \rangle = \sigma_0 \, U$,
and the anomalous terms become
\begin{equation}
\sim  \nf \, \nc  \, \sum_j 
\widetilde{{\cal F}}_j\left(\Phi^\dagger \Phi\right) \; |\det \Phi|^{j/\nc} 
\sim \frac{\nf}{\nc} \, a \, |\log \det U|^2 \; \; , 
\end{equation}
where 
\begin{equation}
a = 
\sum_j  j^2 \, 
\widetilde{{\cal F}}_j\left(\sigma_0^{2 \nf}\right) 
 \; .
\end{equation}
I take ${\cal F}_j = \nf \nc \,\widetilde{{\cal F}}_j$,
with the $\widetilde{{\cal F}}_j \sim O(1)$, which is the expected scaling in $\nf$ and $\nc$ for quarks in the
fundamental representation.
Assuming that the sum over $j$
does not grow with $\nc$, this then 
reduces to the usual Witten-Veneziano term of 
Eq. \eqref{witten-venez} at small $\nf/\nc$. 

Unlike in the 't Hooft limit, in the Veneziano limit there is 
no direct relation between the functions ${\cal F}_j(\Phi^\dagger \Phi)$ 
and the topological
susceptibility of the pure gauge theory.  In the 't Hooft limit,
small $\nf/\nc$ is obviously close to $\nf/\nc = 0$.  
In the Veneziano limit, the topological
susceptibility surely is a nontrivial function of $\nf/\nc$.
The variation of the topological susceptibility with $\nf/\nc$
can be studied by constructing an effective Lagrangian which couples glueballs to 
anomalous terms \cite{DiVecchia:1980yfw,Schechter:1980ak,Giacosa:2026}.

There is a strong argument which justifies Eq. \eqref{general_fracQ},
and which is why I introduced the Veneziano limit of large $\nc$.
I only need one mild assumption, which is merely that as in the 't Hooft limit,
Eq. \eqref{theta_nc_tHooft}, in the Veneziano limit the
free energy is a function not of $\theta$, but of $\theta/\nc$:
\begin{equation}
E(\theta,\nc,\nf) - E(0,\nc,\nf) = a(\nc,\nf) \; \widetilde{E}_{Ve} \left( \frac{\theta}{\nc} \right)
\left( 1 + O\left(\frac{1}{\nc,\nf}\right) \right) \; \; ; \; \;
T < \tchi \; ,
\label{theta_nc_veneziano}
\end{equation}
where
\begin{equation}
a(\nc, \nf) = a_1 \nc^2 
+ a_2 \nc \nf + a_3 \nf^2 \;\; ; \;\; \nc, \nf \rightarrow \infty \; .
\end{equation}
The exact values of the coefficients $a_1$, $a_2$, and $a_3$ are irrelevant.
The point is that the function $\widetilde{E}_{Ve}$, which is of course
different from $\widetilde{E}_{tH}$, is a function only of $\theta/\nc$.

This is most reasonable.  It is impossible to imagine how one could 
continuously extrapolate from small to large
values of $\nf/\nc$
unless this holds.  

The $\theta$ angle appears in the chiral effective Lagrangian as
\begin{equation}
\Phi \rightarrow {\rm e}^{i \theta} \Phi \; .
\end{equation}
The only quantity invariant under $SU(\nf)_L \times SU(\nf)_R$
rotations, but not $U(1)_A$, are powers of $\det \Phi$.
If the effective Lagrangian for $T < \tchi$ is a function of
$\theta/\nc$, then, it is natural for fractional powers of $\det \Phi$,
$\sim (\det \Phi)^{1/\nc}$, to enter. 

While I define the $\eta'$
meson through $\Phi$, as what is apparently a linear field, it really is
not.  Instead, in order to be able to rotate $\theta$ away for massless
quarks, the $\eta'$ field is always an angular variable.  

As discussed by Witten 
at small $\nf/\nc$ \cite{Witten:1979vv,Witten:1980sp}, 
when $\nf/\nc$ is of order one, as in QCD, the effective Lagrangian 
must be a function of $\theta/\nc$, and so multi-valued, 
to ensure that there is a phase transition at $\theta = \pi$
\cite{Gaiotto:2017tne}.  This can
be computed explicitly in this syncretic model \cite{Giacosa:2026}.

The above is only true in the phase where chiral symmetry
is spontaneously broken.  When $T > \tchi$, the free energy is not
a function of $\theta/\nc$, but just of $\theta$.  Thus there is no phase transition at $\theta = \pi$.

I end this section with several comments.  My argument
is based upon the transformation of the effective Lagrangian
under shifts in $\theta$, which 
suggests that on a femto-slab \cite{Cherman:2016hcd,Tanizaki:2022ngt} 
the chiral effective Lagrangian can be written in a form following
Eq. \eqref{general_fracQ}.

In principle, Eq. \eqref{general_fracQ} could be derived from holography
\cite{Arean:2016hcs,Bigazzi:2019hav,Leutgeb:2019lqu,Bonanno:2026zzf}.
This is challenging in the Veneziano limit, as the backreaction of the
quarks on the gluon fields has to be taken into account.  
Ref. \cite{Arean:2016hcs} asserts that the Witten-Veneziano Lagrangian
is unchanged as $\nf/\nc$ changes from small to large values.  I demur.

The above form in Eq. \eqref{general_fracQ} is only valid for
degenerate flavors.  The case of non-degenerate flavors 
is not a trivial extension, and will be treated separately \cite{Giacosa:2026}.

Lastly, it would be most useful to analyze how the ${\cal F}_j(\Phi^\dagger \Phi)$
evolve using the functional renormalization group \cite{Fejos:2026}.  

\section{Direct measurement of fractional 't Hooft
determinants}
\label{sec:direct}

At the outset, I insist that to understand topological charge
on the lattice, it is imperative to have a definition which is explicitly
topological at {\it non}-zero lattice spacing.  There is work by
L\"uscher and Seiberg towards this goal \cite{Luscher:1981zq,Seiberg:1984id},
while J.-Y. Chen has proposed a program using category theory
\cite{Chen:2024ddr,Zhang:2024sgm}.  That this is possible is clear
from the examples involving the first and second homotopy groups
\cite{Chen:2024ddr}.  For example, for the first homotopy group,
this is just Villianization.  It is far less trivial for the third homotopy
group, which is what enters for instanton number.

With trivial boundary conditions, the topological charge is integral.
To see fractional instantons it is necessary to take boundary conditions which are twisted by $Z(\nc)$ gauge transformations.  
I begin by reviewing the discussion of Ref. \cite{Nair:2022yqi},
and then discuss how it can be extended to include dynamical quarks.

For simplicity I take boundary conditions which are twisted at spatial infinity.
In practice, on a lattice instead one will take boundary conditions which are
twisted on a torus, but that is a trivial extension.
For the time like component of a $SU(\nc)$ gauge field,
$A_0$, at spatial infinity I assume 
\begin{equation}
A_0(\vec{x}\,) \; \rightarrow
\; \frac{2 \pi T}{\nc} \; \mathbf{k} \;\; , \;\; r \rightarrow \infty
\;\; , \;\;
\mathbf{k} \; =\; - \; 
\left(
\begin{array}{cc}
{\mathbf 1}_{\nc -1} & 0      \\
0             & -(\nc -1) \\
\end{array}
\right) \; ,
\end{equation}
At $r = \infty$ the Wilson line in the imaginary
time direction, $\tau$, is
\begin{equation}
\Omega \; = \; \exp \left( i \int^{1/T}_0 \; A_0 \; d\tau \right) 
\; = \; \exp\left( \frac{2 \pi i}{\nc} \; \mathbf{k} \right) \; = \;
{\rm e}^{2 \pi i/\nc}\; {\bf 1}_{\nc} \; ,
\end{equation}
so this $A_0$ represents a field with nontrivial $Z(\nc)$ holonomy
at spatial infinity.

For the spatial components of the gauge field 
I construct a split 't Hooft-Polyakov monopole.
Dividing a sphere into its upper and lower hemispheres,
I take different gauge potentials on each patch, as
\begin{equation}
A^\pm_\phi \; = \; \frac{1}{2 \nc \, r } \;
\frac{\left( \pm 1 - \cos\theta \right)}{\sin \theta}  \; \mathbf{m}\; \; , \;\;
\mathbf{m} = \left(
\begin{array}{ccc}
{\mathbf 1}_{\nc-2} & 0      & 0\\
0             & -(\nc-1) & 0 \\
0             & 0      & 1 \\
\end{array}
\right) \; .
\label{zn_soln}
\end{equation}
This represents a $Z(\nc)$ magnetic monopole.  This is shown
by computing the
Wilson line for a special closed path, $\vec{s}$.
Since the vector potential is specified by two patches, we compute
the Wilson line with $A^+$,
going around by $2 \pi$ in $\phi$; and the Wilson line
with $A^-$, running in the opposite direction:
\begin{equation}
\exp \left( \; i \oint \vec{A}^+ \cdot d\vec{s} \; \right)
\; \left( \exp\left( \;
i \oint \vec{A}^- \cdot d\vec{s} \; \right) \right)^\dagger 
 =\exp \left( \frac{2 \pi i }{\nc} \, \mathbf{m} \right)  \;\; = \;\;
 {\rm e}^{2 \pi i/\nc} {\bf 1}_{\nc} \; ,
\end{equation}
which shows that this is a $Z(\nc)$ magnetic monopole.  

The above are the boundary conditions at spatial infinity, $r \rightarrow
\infty$.  At the origin, $r = 0$, we require all $A_\mu$'s to vanish,
at least like $\sim r^2$, so that $F_{\mu \nu} \sim r$ as $r \rightarrow 0$.

It is straightforward to compute the topological charge.  For large $r$,
\begin{equation}
A_0(r) \; = \;
\frac{2 \pi T}{\nc} \; \mathbf{k} \; - \; \frac{1}{2 \nc r} \; \mathbf{m} \; + \; \ldots
\end{equation}
For a static configuration,
\begin{equation}
\topq \; = \;
\frac{1}{4 \pi^2} \int d^4 x \; \partial_i \; {\rm tr} \left( A_0 \; B_i
\right) = \frac{1}{\nc^2} \; {\bf m \cdot k }  = \frac{1}{\nc}\; .
\end{equation}
as first derived by 't Hooft \cite{tHooft:1980kjq}.

This shows how fractional instantons can arise.  Demonstrating
that the confined vacuum is best described by objects with such is
not straightforward.  
Bonanno, D'Elia, and Verzichelli \cite{Bonanno:2020hht}
measured the quantity $b_2$ in Eq. \eqref{top_suscep_conf}
for $\nc = 3$, $4$, and
$6$.  A dilute
gas of instantons with integral charge gives
$b_2 = -1/12$, independent of $\nc$, while a dilute gas of fractional instantons,
with $\topq$ an integer times $1/\nc$, gives
$b_2 = (-1/12)/\nc^2 \approx -0.08/\nc^2$.  Ref. \cite{Bonanno:2020hht} finds that 
$b_2 \approx -0.19/\nc^2$.  
This indicates that the vacuum is composed of fractional
instantons, but since $0.19$ is more than twice $0.08$, it is clearly a {\it dense} liquid of
fractional instantons.  This is reasonable,
as for any $\nc$ one would hardly expect the vacuum
of a $SU(\nc)$ gauge theory is dilute in anything.
For a related study, see Mickley, Kamleh, and Leinweber \cite{Mickley:2023exg}.

With dynamical quarks, as with gluons, one cannot directly detect 
fractional modes without untwisted boundary conditions; 
see, {\it e.g.}, Ref. \cite{Meng:2023nxf,Alexandru:2024tel}.  
For the boundary conditions of quarks, one possibility is to
add $\nf = \nc$ flavors, and adjust a flavor transformation
so that a $Z(\nc)$ transformation in color is compensated
by a $Z(\nf)$ transformation in flavor.  Thus per flavor,
the change in topological charge is $\Delta \topq = 1/\nc$,
so for all flavors, $\nf \Delta \topq = \nf/\nc =1$.

\section{One to four flavors}
\label{sec:one_to_four_flavors}

The form of the fractional anomalous determinants is sensitive to
the number of colors and flavors, so next I consider the syncretic model for three colors and one to four flavors.

Previously, Rennecke and I suggested that
the axial $U(1)_A$ symmetry is dynamically restored
near $\tchi$, in the universality class of 
$SU(\nf)_L \times SU(\nf)_R \times U(1)_A$
\cite{Pisarski:2024esv}.  
For $\nf = 2$ and $4$, the present analysis gives a universality
class of $SU(\nf)_L \times SU(\nf)_R$, but beyond Landau.

\subsection{One flavor}
\label{sec:one_flavor}

The case of one flavor is illustrative, as it is contrary to accepted wisdom.  The chiral field $\Phi$ is just a complex number, so there
are no matrix traces.

If ordinary 't Hooft instantons dominate, the effective Lagrangian is the sum of two pieces \cite{Pisarski:2024esv}.  The first is invariant  under $U(1)_A$ rotations,
\begin{equation}
  \mathcal{L}_{U(1)_A \; {\rm sym}} = |\partial_\mu \Phi|^2 + 
  m^2 \, |\Phi|^2  +
  \lambda \, (|\Phi|^2)^2 \; ,
  \label{one_flavor_sym}
\end{equation}
while the second is not,
\begin{eqnarray}
  \mathcal{L}_{{\rm int.}\, \topq}  & = 
  \xi_1 \, (\Phi + \Phi^\dagger)  + \xi_1^{1} \, |\Phi|^2 \, (\Phi + \Phi^\dagger) + \xi_2 \, (\Phi^2 + (\Phi^\dagger)^2)
  + \xi_2^{1} \, |\Phi|^2 (\Phi^2 + (\Phi^\dagger)^2)  \nonumber \\
  & + \xi_3 (\Phi^3 + (\Phi^\dagger)^3) + \xi_3^1 \, (\Phi + \Phi^\dagger) (\Phi^2 + (\Phi^\dagger)^2) + \xi_4 \, (\Phi^4 + (\Phi^\dagger)^4) + \ldots \; .
  \label{one_flavor_anom}
\end{eqnarray}
The first term in this potential directly generates a mass term for quarks due to the zero mode of an instanton with $\topq = 1$.  
For the $\xi_j^i$, $i$ denotes different couplings with the same
$\topq = j$.

If the anomalous determinants with integral powers
determine the anomalous terms in the effective Lagrangian, Eq. \eqref{one_flavor_anom}, no chiral phase transition occurs.
Instead, as the instantons evaporate, the chiral condensate
just decreases smoothly as $T$ increases.

The syncretic anomalous terms for one flavor and three colors start as
\begin{equation}
{\cal L}_{\rm anom} = m_1^2\, (\Phi^\dagger \Phi)^{5/6}
(\Phi^{1/3} + (\Phi^\dagger)^{1/3} )
+ m_2^2 \,(\Phi^\dagger \Phi)^{2/3}
(\Phi^{2/3} + (\Phi^\dagger)^{2/3} )
+ \xi_1 (\Phi + \Phi^\dagger) + \ldots \; .
\label{frac_one_flavor}
\end{equation}
Obviously there are {\it many} more terms.

To understand the physics in the broken phase, take
\begin{equation}
\Phi = \sigma + i \eta' = \sqrt{\sigma^2 + (\eta')^2} \; {\rm e}^{i 
\theta} \; ; \; \theta = \tan^{-1}\left(\frac{\eta'}{\sigma} \right)\; .
\label{parametrize_eta}
\end{equation}
The first two terms in Eq. \eqref{frac_one_flavor} becomes
\begin{equation}
(\sigma^2 + (\eta')^2 ) 
\left( m_1^2 \;  \cos\left(\frac{\theta}{3}\right)
+ m_2^2\; \cos\left(\frac{2 \theta}{3} \right) \right)\; .
\label{exp_phi_one_flavor}
\end{equation}
The $\sigma$ mass is given by $\theta = 0$, 
\begin{equation}
m_\sigma^2 = m^2 + m_1^2 + m_2^2 \; ,
\label{sigma_mass_one_flavor}
\end{equation}
where I include the mass from $m^2 |\Phi|^2$ in Eq.
\eqref{one_flavor_sym}.  
Expanding in $\eta'$, with $\theta \approx \eta'/\sigma$, the mass for
the $\eta'$ differs, 
\begin{equation}
m^2_{\eta'} = m^2 - \, \frac{1}{18} \left(  m^2_1 + 4 \, m^2_2 \; \right) \; .
\label{eta_mass_one_flavor}
\end{equation}
%
Whether the $\sigma$ or the $\eta'$ is the lightest field depends upon the sign
of $m_1^2$ and $m_2^2$.  

Expanded naively, Eqs. \eqref{parametrize_eta} and \eqref{exp_phi_one_flavor} 
also appear to indicate that there are terms $\sim (\eta')^4/\sigma^2$, which
would be singular as $\sigma \rightarrow 0$.  If one takes the limit before
expanding, though, 
$\theta \rightarrow \pi/2$, and the quadratic terms are well behaved.
This does illustrate that fractional powers of the anomalous
determinant are ill-defined in a symmetric phase, and is why the
present model must be syncretic.

The usual assumption about one flavor is that $\xi_1$ is large,
so there is no trace of even an approximate chiral symmetry in the vacuum.
A syncretic  model suggests an alternate picture.  The coupling $\xi_1$ is small
in vacuum, and chiral symmetry breaking is driven by a large,
negative value for $m^2$, as is usual in gauge theories.

Even if the coupling $\xi_1$ is small, I see no reason why
it should vanish.  If $\xi_1 \neq 0$, the chiral transition
is crossover. Sill,  for small $\xi_1$, there could
well be a significant peak in the susceptibility for the chiral condensate.
Since the $\eta'$ is always massive, the associated universality
class is $Z(2)$, as only the $\sigma$ meson is
nearly critical.  

Computing the behavior
as $T \rightarrow \tchi^+$ with Eq. \eqref{one_flavor_anom} is trivial,
but how to treat that as $T \rightarrow \tchi^-$ with Eq. \eqref{general_fracQ}
is not.  Certainly the chiral
phase transition is beyond Landau. 

The case of one flavor is also illuminating in that it clearly shows why
non-polynomial terms {\it must} arise in the chiral effective Lagrangian.
Without such terms, the $\eta'$ would be like the $\sigma$, and massless
at the putative $\tch$.  It is only because of the terms $\sim (\det \Phi)^{1/\nc}$
that the $\eta'$ can remain massive at $\tchi$.

In essence, unlike the $\sigma$ field, the $\eta'$ is necessarily an angular
variable.  This is obvious from the fact that to eliminate the $\theta$ angle
for massless quarks requires a rotation in the direction of the $\eta'$.

Because the prefactors for
fractional anomalous determinants were chosen
to ensure that they have the
same mass dimension as $\Phi^\dagger \Phi$ as $\Phi \rightarrow 0$,
the example above illustrates that because fractional powers of
$\det \Phi$ enters, the masses nevertheless exhibit non-analytic behavior.
Indeed, the difference of the $\sigma$ and $\eta'$ masses above
is crucial to ensuring that the $\eta'$ is massive at $\tch$.

I conclude this section with a comment.  
For one flavor the field
$\Phi$ can be redefined as
\begin{equation}
\Phi(x) = \phi(x)^{3} \; .
\end{equation}
In terms of $\phi(x)$, fractional powers of the anomalous determinant are then
integral.  However, the total effective Lagrangian is then
\begin{equation}
{\cal L} = 4 |\phi^{2} \partial \phi|^2 + m^2 (|\phi|^2)^3 + \lambda 
(|\phi|^{2})^6
+ \sum_{j=1}^{\infty} {\cal F}_j\left((|\phi|^2)^3\right) \;  \left(\phi^{j} + (\phi^*)^{j} \right) \; .
\label{one_flavor_nonpoly}
\end{equation}
This is certainly a non-standard Lagrangian in field theory, especially given the peculiar powers of $\phi$ in the
kinetic term.  For this reason I expect the fractional powers
of the anomalous determinant only appear in the phase with broken
chiral symmetry.

The case of one flavor provides the
most dramatic difference in distinguishing between a
theory dominated by fractional, as opposed to
integral, topological charge.  On the lattice, however,
with a single flavor it is challenging measuring masses, because disconnected diagrams contribute.

Thus I turn to two flavors, where lattice measurements
are far simpler.

\subsection{Two flavors}
\label{sec:two_flavors}

For two symmetric flavors and three colors, in a syncretic model
the anomalous terms start as
\begin{align}
{\cal L}_{\rm anom} & = 
m_1^2 \left( \det \Phi^\dagger \Phi\right)^{1/3} \,
\left( (\det \Phi)^{1/3} + (\det \Phi^\dagger)^{1/3} \right)
\nonumber \\
& + m_2^2 \left(\det \Phi^\dagger \Phi\right)^{1/6}
\left( (\det \Phi)^{2/3} + (\det \Phi^\dagger)^{2/3} \right)\nonumber \\
& + \xi_1 \, \left( \det \Phi + \det \Phi^\dagger \right) + \ldots
\label{anom_nf2_nc3}
\end{align}
I have assumed that the ${\cal F}_1$ and ${\cal F}_2$
are functions only of $\det \Phi^\dagger \Phi$, and 
not of other invariants, such as ${\rm tr} \, \Phi^\dagger \Phi$,
and ${\rm tr} \, (\Phi^\dagger \Phi)^2$.  An analysis using the functional renormalization
group indicates that the complete form does involve all of these invariants
\cite{Fejos:2026}.
Thus my analysis here is only meant to be illustrative.

Making this assumption, the prefactors of $\det \Phi^\dagger \Phi$ in the first
two terms, with powers of $\frac{1}{3}$ and $\frac{2}{3}$,
are chosen
so that they have dimensions of $\sim \Phi^2$ as $\Phi \rightarrow 0$.
For two flavors, $\det \Phi$ also has dimensions of $\Phi^2$.

The elements of $\Phi$ are
\begin{equation}
\Phi = (\sigma + i \, \eta') \, t^0
+ (a_0^a + i \, \pi^a)\, t^a \; . \;
\label{definition_Phi}
\end{equation}
For two flavors the $\eta'$ is actually the $\eta$ meson;
similarly, for $\nf =3$ the $\pi^a$ include pions,
kaons and the $\eta$ meson,
{\it etc.}  I keep the above notation because it is cleaner for all
$\nf$.  The $\sigma$ and the $a_0^a$ have $J^P =0^+$, and the
$\eta'$ and the $\pi^a$, $J^P = 0^-$.  

With this assumption, both $m_1^2$, $m_2^2$, and $\xi_1$
split the $\eta'$ from the $\sigma$.  Thus if
the chiral transition is of second order (it is, unless a quartic
coupling is negative), then the universality class is $O(4)$, as
usual without fractional anomalous determinants.

The anomalous terms also split the isotriplet $\pi^a$ from the $a_0^a$,
which provides a way to measure $m_1^2$, $m_2^2$, and $\xi_1$.  
Since disconnected diagrams do not contribute to the isotriplet
channels, this splitting is much easier to measure on the lattice.  I explain this more fully in the next subsection for $2+1$ flavors,
but to summarize: the mass difference between the $\pi^a$ and the
$a_0^a$ can be measured on the lattice.
In practice, it is easier to measure the susceptibilities, Eq. \eqref{pi_a_splitting}.  One then need to compute
the (renormalized) chiral condensate, 
$\langle \Phi \rangle = \sigma_0 \, {\bf 1}_{\nf}$,
in the chiral limit, as $T \rightarrow \tchi^-$.  As $\sigma_0 \rightarrow 0$, the anomalous terms scale as
\begin{equation}
\sim m_1^2 \,  \sigma_0^{4/3} \; \; ; \;\; 
\sim m_2^2\,  \sigma_0^{2/3}\; \; ; \;\; 
\xi_1 \; .
\end{equation}
Note that for two flavors, $\xi_1 \, \det \Phi$ is a mass term,
and so it is constant as $\sigma_0 \rightarrow 0$.
Away from the strict chiral limit the transition is 
crossover, with $\tchi$ fixed by the peak
in the chiral susceptibility.  Then it is best to
the mass (or susceptibility) difference around $\tchi$.

\subsection{Three flavors}
\label{sec:three_flavors}

For three symmetric flavors and three colors, in a syncretic model
the anomalous terms start as
\begin{align}
{\cal L}_{\rm anom} & = 
m_1^2 \, \left( \det \Phi^\dagger \Phi\right)^{1/6} \,
\left( (\det \Phi)^{1/3} + (\det \Phi^\dagger)^{1/3} \right)
\nonumber \\
& + m_2^2 \, 
\left( (\det \Phi)^{2/3} + (\det \Phi^\dagger)^{2/3} \right)\nonumber \\
& + \xi_1 \, \left( \det \Phi + \det \Phi^\dagger \right) + \ldots
\label{anom_nf3_nc3}
\end{align}
In this case, $m_1^2$ and $m_2^2$ both contribute to the mass of
the $\eta'$, while the term $\sim \xi_1 \, \det \Phi$ is cubic in
$\Phi$.  

As for two flavors, to illustrate the possible behavior of the model
I assume that ${\cal F}_1$ only involves
$\det \Phi^\dagger \Phi$.  The functional renormalization group 
indicates that the complete form involves
more invariants of $\Phi^\dagger \Phi$ \cite{Fejos:2026}.

In the standard model for the chiral phase transition,
Eq. \eqref{eq:eff_lag_ua1_vio} applies at all temperature.
For the $\eta'$ to be heavy at zero temperature, the
coefficient of the term with $\topq = 1$, $\xi_1$,
must be large \cite{Pisarski:2016ixt}.  It is so large,
in fact, that the $m^2$ in Eq. \eqref{eq:eff_lag_ua1_sym} is positive,
and not negative, as one would expect with the spontaneous
breaking of chiral symmetry in vacuum \cite{Pisarski:2016ixt}.

In contrast, in the syncretic model the fractional 't Hooft
determinants $\sim m_1^2 \, (\det \Phi)^{1/3}$
and $\sim m_2^2 \, (\det \Phi)^{2/3}$ contribute below
$\tchi$, but not above.  Thus the mass of the $\eta'$ at zero temperature is determined not just by $\xi_1 \sigma_0$, but also by
$m_1^2$ and $m_2^2$.  Thus the value of the
term with $\topq = 1$, $\xi_1$, need {\it not} be large.

This affects the order of the chiral transition.
Since there is always a cubic term 
$\sim \det \Phi$ from $\topq = 1$, the transition remains of first
order  \cite{Pisarski:1983ms}.  Even so, if $\xi_1$ is small, with the mass of the
$\eta'$ due largely to $m_1^2$ and $m_2^2$,
the chiral transition could well be {\it weakly} first order, and possibly very weakly.  This would explain the lattice
results
\cite{Cuteri:2021ikv,Dini:2021hug,DAmbrosio:2025imc,Zhang:2026zpa,Ding:2026blh}, which as of yet see no sign for a first order transition.

There are recent results by Zhang {\it et al.} \cite{Zhang:2026zpa}, who studied three degenerate flavors with Mobi\"us fermions. At the lightest masses, their Fig. (14) indicates surprising results for the height of the total chiral susceptibility. If $N_s$ is the number of spatial lattice spacings, the peak height scales as $N_s^\alpha$. For a first order transition, $\alpha = 3$, while $\alpha \approx 1.96$ for a second order transition in the universality class of $Z(2)$. Instead, they measure $\alpha \approx 0.39 \pm 0.04$ -- a scaling too weak for either -- and together with the fact that this finite-volume effect can be well described by an exponential fit (although larger-volume simulations are needed to confirm the saturation), they suggest that the transition is a crossover.  Instead, it may signal that the chiral transition is beyond Landau
\cite{Chen:2025generalized}. 

QCD involves two light quarks and a heavier strange quark.  The extension
of my results to non-degenerate flavors is not obvious, and will
be analyzed in another paper \cite{Giacosa:2026}.
Even so, it is possible to make a few general remarks to motivate
how to measure the parameters $m_1^2$, $m_2^2$,
and $\xi_1$ from numerical simulations of lattice QCD.
The (approximate) restoration of the axial $U(1)_A$ symmetry
has been studied on the lattice by many groups
\cite{Bazavov:2012qja,Buchoff:2013nra,Bhattacharya:2014ara,Aoki:2015qwa,Bonati:2015vqz,Dick:2015twa,Borsanyi:2016ksw,Tomiya:2016jwr,Bonati:2018blm,Kotov:2019dby,Ding:2020xlj,Aoki:2020noz,Borsanyi:2020fev,Kaczmarek:2020sif,Aoki:2021qws,Kaczmarek:2021ser,Lahiri:2021lrk,Borsanyi:2022fub,Kaczmarek:2023bxb,Alexandru:2024tel,Borsanyi:2025lim,Kotov:2025ilm,Aarts:2026kpq,Chiu:2026sxy,Chiu:2026upk}.  The consensus appears to be that the axial $U(1)_A$
is still badly broken by the crossover temperature, $\tchi$.  
At high temperature, when there is a dilute instanton gas,
the topological susceptibility falls like a high power of the temperature.

At what temperature the axial $U(1)_A$ symmetry is approximately restored is still unsettled.  Most groups find a temperature of $\approx 200$~MeV,
while Ref.
\cite{Aarts:2026kpq} find $\approx 300$~MeV.
In the case of two massless flavors, Refs.
\cite{Chiu:2026sxy,Chiu:2026upk} find that
the axial $U(1)_A$ symmetry is (approximately) restored {\it at} $\tchi$.
The latter is in accord with what I suggest.

Clearly it would be useful to have detailed measurements to distinguish
between breaking of the axial $U(1)_A$ symmetry by instantons with
fractional versus integral topological charge.  Most simulations are
done with $2+1$ flavors, with the up and down flavors light, and the strange, not so light.  Since for three flavors $\det \Phi$ is cubic in $\Phi$, any
term involving $\det \Phi$ --- whether with integral 
or fractional powers ---
contributes to the masses in the chirally symmetric phase accompanied
by a single power of a condensate
\cite{Shuryak:1993ee, Cohen:1996ng, Lee:1996zy, Evans:1996wf}.
In QCD, the chiral condensate for the light quarks is
small above $\tchi$, which is why the peak in the 
chiral susceptibility at $\tchi$ is greatest for light quarks.
The chiral condensate for the strange quark decreases
above $\tchi$, but falls off much slower, and 
decreases between $\tchi$ and $2 \, \tchi$.
Thus the most useful signal
is to measure the splitting between the light mesons, the $\pi^a$
and $a_0^a$, as that is multiplied by the strange quark condensate.
The splitting between the analogous strange particles, such as kaons
and their $J^P = 0^+$ partners, is proportional to the condensate
for the light quarks, and thus is much smaller.

The splitting between the $\pi^a$ and the
$a_0^a$ channels is given entirely by the anomalous terms in
the effective Lagrangian, and so is a sum of three contributions,
\begin{equation}
m_{a_0}^2 - m_\pi^2 \approx \; 
m_1^2 \, \sigma_s^{4/3}
\; , \; m_2^2 \, \sigma_s^{2/3}
\; , \; \xi_1 \, \sigma_s \; ,
\label{mass_splitting_nf3}
\end{equation}
where $\sigma_s$ is the chiral condensate for the strange quark.

On the lattice it is easier to measure
the two point function of the correlator at
zero momentum, 
\begin{equation}
\zeta_\pi(T) = \int d^3 x \; \langle \pi^a(x) \pi^a(0) \rangle
\sim \frac{1}{m_\pi(T)^2} \; ,
\label{pi_a_splitting}
\end{equation}
up to a form factor, and similarly for the $a_0^a$, $\zeta_{a_0}$.  
Normalizing by forming the ratio of the difference and sum
\cite{Aarts:2026kpq}, 
\begin{equation}
{\cal R}_{\pi, a_0}(T) = \frac{\zeta_{\pi}(T) - \zeta_{a_0}(T)}{\zeta_{\pi}(T) + \zeta_{a_0}(T)} \; .
\label{diff_masses}
\end{equation}
I suggest fitting to a series in $\sigma_s^{4/3}$, $\sigma_s^{2/3}$,
and $\sigma_s$, following Eq. \eqref{mass_splitting_nf3}.

For $T > 300$~MeV, then, where a dilute instanton gas dominates,
it must be true that $m_1^2$ and $m_2^2$ are near
zero, and Eq. \eqref{diff_masses} is dominated by the term with
$\topq = 1$, $\sim \xi_1 \, \det \Phi$.  
The interesting region is $T: 150 \rightarrow 300$~MeV, 
where $\sigma_s$ decreases significantly.  

This should be an especially sensitive test of my ansatz
for ${\cal F}_1$ and ${\cal F}_2$, which as I emphasized
repeatedly, is rank supposition.  Indeed, that the powers of
$\sigma_s$ are, in order,
$\frac{4}{3}$, $\frac{2}{3}$, and $1$,
Eq. \eqref{mass_splitting_nf3}, seems most odd.  Input from
lattice simulations will really be useful in constructing
a better effective Lagangian and fixing ${\cal F}_1$ and ${\cal F}_2$.

Extracting the value of the chiral condensate for the strange quark is
feasible, but not trivial.  On the lattice, the 
chiral condensates have quadratic and logarithmic ultraviolet divergences.  To eliminate them, what is uniformly measured is
\cite{Cheng:2006qk,Borsanyi:2010bp,Bazavov:2011nk}
\begin{equation}
\Delta_{l,s}(T) = \frac{ \langle \overline{l} l\rangle(T) -
(m_l/m_s) \langle \overline{s} s\rangle(T)}{ \langle \overline{l} l\rangle(0) -
(m_l/m_s) \langle \overline{s} s\rangle(0)} \; ,
\end{equation}
where $l$ are the up and down quarks, with degenerate masses, and $s$ is the
strange quark.
However, this does not indicate the relative magnitude of the two condensates.
Instead, to measure the chiral condensate for the strange quark,
it may be better to measure \cite{Golan:2026}
\begin{equation}
\widetilde{\Delta}_{l,s}(T) = \frac{ \langle \overline{l} l\rangle(T) -
\langle \overline{l} l\rangle(0)}{ \langle \overline{s} s\rangle(T) -
\langle \overline{s} s\rangle(0)} \; .
\end{equation}

The above suggests a cartoon of the phase transition in QCD.  The spontaneous
breaking of chiral symmetry occurs in vacuum, so all quarks are massive.  
Each massive quark binds to a fractional instanton with
charge $\frac{1}{3}$, which surely forms a liquid.
The up and down flavors become light at the chiral crossover
temperature of $\tchi \approx 155 \pm 5$~MeV.  
Above $\tchi$, the (near) zero modes
of light up and down quarks bind two charge $\frac{1}{3}$ fractional instantons
into one with charge $\frac{2}{3}$, while the massive strange quark binds to a fractional instanton
with charge $\frac{1}{3}$.   From the measurements described
above, the chiral condensate for the strange quark vanishes at a higher temperature,
$T_{\chi s}$.  Above this temperature, the up and down quarks are very light,
and the strange quark has a current quark mass of $\approx 100$~MeV.  
It is possible that this asymmetric pairing, of up and down massless quarks
into fractional instantons with charge $\frac{2}{3}$, and a massive strange
quark with charge $\frac{1}{3}$, persists above $T_{\chi s}$.  Of course
at high temperature, where the strange quark mass can be neglected,
the massless up, down, and strange quarks bind fractional instantons
with charge $\frac{1}{3}$ into instantons with integral charge.

\subsection{Four flavors}
\label{sec:four_flavors}

For four symmetric flavors and three colors, $\det \Phi \sim \Phi^4$,
so the syncretic anomalous terms begin as
\begin{align}
{\cal L}_{\rm anom} & = 
m_1^2 \; \left( \det \Phi^\dagger \Phi\right)^{1/12} \,
\left( (\det \Phi)^{1/3} + (\det \Phi^\dagger)^{1/3} \right)
\nonumber \\
& + \; m_2^2 \; (\det\Phi^\dagger \Phi)^{\alpha_2}\,
\left( (\det \Phi)^{2/3} + (\det \Phi^\dagger)^{2/3} \right)\nonumber \\
& + \xi_1 \, \left( \det \Phi + \det \Phi^\dagger \right) + \ldots
\label{anom_nf3_nc3}
\end{align}
Again, I assume that ${\cal F}_1$ and ${\cal F}_2$
are functions only of $\det \Phi^\dagger \Phi$.  The weakness of
this assumption is particularly clear as one increases the number of flavors.

For example, assuming that the ${\cal F}_2$ is a function only of
$\det \Phi^\dagger \Phi$, the value of the coefficient $\alpha_2$ is
unclear.  If one requires
that the product of $(\det\Phi^\dagger \Phi)^{\alpha_2} \, (\det \Phi)^{2/3}$
reduces to $\sim \Phi^3$ as $\Phi \rightarrow 0$, then $\alpha_2 = 1/12$; 
if it reduces to
$\sim \Phi^4$, then $\alpha_2 = 1/6$.  Certainly the assumption that it
reduces to $\sim \Phi^2$ is too extreme.  

Thus as found for two flavors with the functional renormalization group
\cite{Fejos:2026}, the functions ${\cal F}_1$ and ${\cal F}_2$ will surely
involve not just $\det \Phi^\dagger \Phi$, but also fractional powers
of ${\rm tr}\, \Phi^\dagger \Phi$, ${\rm tr} (\Phi^\dagger \Phi)^2$, {\it etc.}.
Presumably this reflects the fact that as $\nf$ increases,
the zero modes of massless quarks pair fractional instantons into
integral instantons in more complicated ways.

If $\alpha_2 = 1/12$ there is an interesting possibility.  
Halperin, Lubensky, and Ma
showed that for an Abelian Higgs model, in three dimensions
the presumptive second order phase transition is instead
of first order, induced by a cubic
term $\sim - (|\phi|^2)^{3/2}$ \cite{Halperin:1973jh}.  If this were just an ordinary
cubic term, $\sim \phi^3$, then the transition would inevitably
be first order.  Since $|\phi|^2$ enters, though, 
the sign of the coupling matters.  For positive coupling
the transition is crossover, although of an unusual form.
To one loop order, though, the coupling is negative,
and the transition is weakly first order.  Hence
for four flavors the term with $\topq = \frac{2}{3}$ could drive the chiral
transition first order, although surely it is weakly so.

In any case, for four flavors the fractional anomalous determinant from
$\topq = \frac{2}{3}$ is more important for the chiral phase transition than that from $\topq = \frac{1}{3}$.  This is unlike $\nf < 4$.

The anomalous determinant from $\topq = 1$, $\sim \det \Phi$, is of quartic
order, and so a marginal coupling in four spacetime dimensions.
If only integral anomalous determinants enter about $\tchi$,
then the axial $U(1)_A$ symmetry is not restored at $\tchi$,
and the critical behavior is determined by three quartic couplings,
for $(\trace \Phi^\dagger \Phi)^2$,
$\trace(\Phi^\dagger \Phi)^2$, and $\det \Phi$.  Even so,
as $T \rightarrow \tchi$, the masses for the $\sigma$ and $\eta'$
are the same; the breaking of $U(1)_A$ is only seen in the 
anomalous determinant with $\topq = 1$, $\sim \det \Phi$.

As for two and three flavors, 
by measuring the splitting of the $\pi^a$ and $a_0^a$ fields,
and the expectation value of the chiral condensate, 
the dependence of ${\cal F}_1$ and ${\cal F}_2$
on $\sigma_0$ can be fixed. Since four flavors does not require rooting,
this is a particularly valuable example.

\section{Baryon number in the symmetric phase}
\label{sec:baryon}

In the nonlinear sigma model, besides the standard terms of
Eq. \eqref{nonlinear_L}, there is also a topological term,
\begin{equation}
n_B(U) = \frac{1}{24 \, \pi^2} \int d^3 x \,
\epsilon^{i j k} \, \trace \left( V_i \, V_j \, V_k \right)
\; , \; V_i = U^\dagger \, \partial_i \, U \; .
\label{baryon_number_nonlinear}
\end{equation}
Here $V_i = U^\dagger \partial_i U$ looks like a
pure gauge transformation in a $U(\nf)$ gauge theory.
As such, the associated field strength vanishes, and
this is why $n_B(U)$ must be an integer. 
This is because in topology, 
the third homotopy group is given by the integers,
$\pi_3(SU(\nf)) = {\cal Z}$ \cite{Zumino:1983rz}.
Similarly, in the Skyrme model 
$n_B$ is the baryon number \cite{Witten:1983tw}.

This term is essential to understand the structure of 
the 't Hooft anomaly for the chiral phase transition.
As shown by CCP \cite{Chen:2026jla}, if one adds 
a term for an imaginary chemical potential to the Lagrangian,
$\sim i \theta_B n_B$, in the plane of
$T$ and $\theta_B$ it is most likely that the chiral phase
transition has a critical line in $\theta_B$,  which is beyond  Landau \cite{Chen:2025generalized}.  

Thus there must be an analogous term in the 
linear sigma model.  A first guess is simply to replace $U$ by $\Phi$ in $V_i$, taking $V'_i = \Phi^\dagger \partial_i \Phi$.  This doesn't work because $V'_i$, treated as the vector potential,
is not a pure gauge transformation.  Thus the corresponding field strength doesn't vanish, and $n_B$ varies continuously.

For a single flavor, $\Phi$ is a complex number, and a phase
is given just by taking $\Phi/|\Phi|$.  (For a single
flavor, $n_B$ is not an integer, as $\pi_3(U(1))$ is trivial.)  
Taking care with ordering, this can still be done when $\Phi$ is a matrix,
\begin{equation}
\widetilde{U} = \Phi \; \frac{1}{\sqrt{\Phi^\dagger \Phi}} \; ; \;
\det \Phi \neq 0 \; .
\label{eq:linear_to_nonlinear}
\end{equation}
This yields a unitary
matrix $\widetilde{U}$ if and only if $\Phi$ is invertible.
That is, if $\det \Phi$ is nonzero.

From $\widetilde{U}$ I then define
$\widetilde{V}_i = \widetilde{U}^\dagger \partial_i
\widetilde{U}$, and compute $n_B(\widetilde{U})$.
Since by construction $\widetilde{U}$ is unitary,
$n_B(\widetilde{U})$ is automatically an integer.  
From topology, this reflects the fact that   
$\Phi$ is an element of $GL(\nf,{\cal C})$, so with this decompostion, $\pi_3(GL(\nf,{\cal C})) = \pi_3(SU(\nf)) = {\cal Z}$.  Similar
considerations also arise
for Chern-Simons theories with a complex gauge
group \cite{Witten:1989ip,Witten:2010cx}.  

This leaves open the question of what happens when
$\det \Phi$ vanishes and $\Phi$ is not invertible.
In space of configurations in $\Phi$, when
$\det \Phi$ goes to zero the baryon number can jump.
But these are clearly isolated, special points.
Modulo this subtlety, this explains how to relate
the results of CCP in the nonlinear model \cite{Chen:2026jla} 
to the linear model used here.  Admittedly, given
the expression for $\widetilde{U}$, the operator
for baryon number is non-polynomial in $\Phi$, but then again,
so are the anomalous terms in the broken phase.  
The connection between the baryon number and
the anomalous determinant is intriguing.

\section{Conclusions}
\label{sec:conclusions}

I end by emphasizing that Eq. \eqref{diff_masses} is directly
testable with current lattice techniques.  
After extending the present results to $2+1$ flavors \cite{Giacosa:2026},
all that is necessary is to
measure the two-point functions of the $\pi^a$ and $a_0^a$ fields
at zero momentum, and the renormalized values of the chiral
condensate for the strange quark, particularly 
between $T \sim 150$ and $300$~MeV.  As was discussed above,
the couplings for fractional anomalous determinants can also be determined
for one to four flavors.  The case of one flavor is especially
interesting, but is challenging numerically.

It is imperative to analyze the case 
of $2+1$ flavors, when one takes either the light, or the strange, quarks
to be massless \cite{Giacosa:2024orp,Giacosa:2026,Kovacs:2026,Debnath:2026}. 

Recently the BESIII collaboration has claimed the discovery of a
pseudoscalar glueball, the X(2370) \cite{BESIII:2023wfi,BESIII:2026mvn}.
In a related paper, Giacosa, Jafarzade, Kov{\'a}cs, Kov{\'a}cs, Rennecke
and I analyze the effects which fractional anomalous determinants
have on the decays of the X(2370) \cite{Giacosa:2026}.  We also
analyze the behavior of the $\theta$-vacua in QCD, and estimate the
strength of the first order transition at $\theta = \pi$.

A serious shortcoming of my analysis is any understanding of the form of
the functions ${\cal F}_j(\Phi^\dagger \Phi)$.  Fejos is analyzing this
using the functional renormalization group \cite{Fejos:2026}.

The appearance of fractional anomalous determinants has strong
implications for the phase diagram of QCD in the plane of temperature,
$T$, and chemical potential, $\mu$ \cite{Nogradi:2026}.  In particular, there is suprising behavior
for the anomalous $U(1)_A$ symmetry
in the chirally symmetric phase at zero temperature  \cite{AragaodeCarvalho:1979xu,AragaodeCarvalho:1980de,Bruckmann:2013rpa,Nogradi:2026}.

Previously, a chiral matrix model was developed, in which a matrix
model for "deconfinement" is married to a quark-meson model
\cite{Pisarski:2016ixt}.  In this model, only the integral
anomalous determinant with $\topq = 1$ was used, $\sim \xi_1 \det \Phi$.  Because the $\eta'$
is so heavy, the coupling $\xi_1$ is large,
so that in vacuum, $m^2 > 0$.  Consequently,
there is no critical endpoint in the $T-\mu$ plane.  Adding fractional anomalous determinants radically changes the analysis, as $m_1^2$ and $m_2^2$ can be 
used to make the $\eta'$ heavy \cite{Debnath:2025qhd,Debnath:2026}, with
a small value of $\xi_1$.

There are three other areas where these results might be of use.
The first is for axions in QCD and as a candidate for dark matter
\cite{Rennecke:2020zgb}.
Second, Veneziano noted that the $U(1)_A$ anomaly 
is relevant for the structure function $g_1^p$
in polarized deep inelastic scattering
\cite{Veneziano:1989ei,Shore:2007yn,Tarasov:2020cwl,Tarasov:2021yll,Tarasov:2025mvn,Bhattacharya:2022xxw,Bhattacharya:2023wvy}.  Lastly, 
in condensed matter physics axionic charge density waves in Weyl semi-metals
are analogous to the $\eta'$
\cite{Zyuzin:2012tv,Goswami:2013PRB,Wang:2013PRB,Gooth:2019lmj,Mottola:2023emy}.  In all of these cases, presumably fractional anomalous determinants 
changes the physics.

The example of femto-slabs 
\cite{Cherman:2016hcd,Tanizaki:2022ngt,Hayashi:2024qkm} are
most useful in pointing the way to understanding terms which
break axial $U(1)_A$ in infinite volume.
Especially illuminating is a femto-torus \cite{Hayashi:2024qkm},
which reduces to an effective theory in $1+1$ dimensions.  Unfortunately,
continuous symmetries cannot be broken in two dimensions, and so there
cannot be a true chiral phase transition; 
perhaps there are Berezinskii-Kosterlitz-Thouless phase 
transitions.  A femto-torus could be used to study the
detailed form of the ${\cal F}_j$'s for non-degenerate
flavors, Eq. \eqref{general_fracQ}.  

A closely
related limit is that of a femto-sphere, taking a sphere of radius $R$ so small that the QCD coupling constant is also small
\cite{Sundborg:1999ue,Aharony:2003sx,Aharony:2005bq}.  Integrating out
modes with nonzero angular momentum allows one to compute the effective
Lagrangian for the constant modes.  While there is no true phase transition
at finite volume, there is when the group volume is infinite, taking $\nc \rightarrow \infty$.   There is a 
first order transition for pure glue, which becomes of third order
with dynamical quarks \cite{Schnitzer:2006xz}.  
Presumably it is possible to compute the chiral effective Lagrangian
for massless quarks; $Z(\nc)$ vortices should exist on a femto-sphere,
and the phase transition as $T \rightarrow \tchi$ could be analyzed analytically.

As I was finishing this paper, two related papers appeared.
First, D'Elia, Siliberto, and Zambello computed the
phase diagram for $2+1+1$ flavors in the plane of $T$ and
imaginary chemical potential, $\theta_B$ \cite{DElia:2026ues}.
Along the Roberge-Weiss line at $\theta_B = i \pi/3$, they find
that the coefficient $b_2$ exhibits a sharp jump from fractional
to integral instantons right at $T_{RW}$.  Evidently, at
this $\theta_B$ the nonzero quark masses do not prevent
quarks from binding fractional into integral instantons.
As discussed at the end of Sec. \ref{sec:three_flavors}, 
this is unlike $\theta_B = 0$, where the chiral condensate for
the strange quark affects this binding between $\tchi$ and at least $T_{\chi s}$.

Second, Di Vecchia, Sannino, Shore, Veneziano and Yankielowicz proved a theorem
that solving the CP problem in the vacuum of 
QCD requires either a massless quark, or a
very light $\eta'$ \cite{DiVecchia:2026mrz}.  Their analysis carries over unchanged, with their
Eq. (10) equivalent to my Eq. \eqref{eq:linear_to_nonlinear}.

I conclude by paying homage, once again, to the mysteries of the axial
anomaly \cite{Adler:1969gk,Bell:1969ts}.   Since the breaking of
the $U(1)_A$ symmetry is quantum mechanical, the case of nonzero
temperature and density provides an especially interesting example,
as the breaking is large in vacuum, with a heavy $\eta'$, to its
asymptotic restoration as $T$ and/or $\mu \rightarrow \infty$.
In between, the  chiral phase transition is syncretic,
exclusively because of the anomaly.

\section{Acknowledgments}
This work is a direct outgrowth of work with S. Chen and A. Cherman \cite{Chen:2026jla}.  I am most grateful to them for that collaboration,
and for insisting that any linear sigma model needs to explain our previous work,
which led to Sec. \ref{sec:baryon}.
I also thank: S. Chen, for immediately pointing out that my first guess, Eq. \eqref{first_guess}, is obviously wrong; J. Schaffner-Bielich, who
at the meeting on "Compact Stars in QCD 2026" (Barcelona, May, 2026) asked
me a question that prodded me to think once again
about the generalization of the Witten-Veneziano term to large $\nf/\nc$; 
H. Davoudiasl, J.L.D. Golan, Y. Mehtar-Tani, S. Mukherjee, and P. Petreczky for discussions, especially about lattice QCD; S. Chen, A. Cherman,
T. Dumitrescu, Z. Komargodski, T. Sulejmanpasic, D. Tong, and M. Unsal, for discussions at the 
meeting by the Simons Collaboration on "Paths to Quantum Field Theory", June 22-26, 2026, where 
uniformly all said that when $\nf/\nc$ is large,,
the $\eta'$ is massive at $\tchi$, but they didn't know how to
generalize the Witten-Veneziano term; E. Mottola,
for numerous discussions about the anomaly in June, 2026;
Y. Zhang, for detailed explanations of his work in
Ref. \cite{Zhang:2026zpa}; T.-W. Chiu, for discussions on
his work in Refs. \cite{Chiu:2026sxy,Chiu:2026upk}; the Nuclear
Theory Group at Lawrence Berkeley Laboratory, for a visit in July, 2026,
for discussions with R. Briceno, K.-F. Liu,  A. Walker-Loud, X.-N. Wang, and F. Yuan, and inviting me to
speak about this work; F. Nogueira, Z. Nussinov, M. Ogilvie, and S. Schindler, 
for discussions, and reminding me of
Ref. \cite{Halperin:1973jh}; M. Debnath, R. Ghosh, N. Haque, Y. Hidaka, and 
V. Skokov \cite{Debnath:2026} and also F. Giacosa,  
G. Kovacs, P. Kovacs, and F. Rennecke \cite{Giacosa:2026},
especially about non-degenerate quark masses; A. Rebhan, for
discussions about holographic models; D. Nogradi, for discussions
on instantons at $\mu \neq 0$; G. Fej{\"o}s, for discussions about the
functional renormalization group \cite{Fejos:2026}; 
and R. Paatelainen, for the invitation to present this work
at SEWM 2026, Helsinki, and Jyvaskyla in August, 2026.  RDP is supported by the U.S.
Department of Energy under contract DE-SC0012704, and by the Alexander
von Humboldt Foundation. 
Claude Fable 5 was used extensively in this research, and was crucial in Sec. \ref{sec:baryon}.

%

\end{document}